\documentclass[prb,amsmath,amysymb,amsfonts,superscriptaddress,floatfix,twocolumn]{revtex4-1}
\usepackage{color}
\usepackage{bm}
\usepackage{graphicx}
\usepackage{hyperref}

\begin{document}

\title{Projected Hartree--Fock Theory}

\author{Carlos A. Jim\'enez-Hoyos}
\affiliation{Department of Chemistry, Rice University, Houston, Texas 77005, USA}

\author{Thomas M. Henderson}
\affiliation{Department of Chemistry, Rice University, Houston, Texas 77005, USA}
\affiliation{Department of Physics and Astronomy, Rice University, Houston, Texas 77005, USA}

\author{Takashi Tsuchimochi}
\affiliation{Department of Chemistry, Rice University, Houston, Texas 77005, USA}

\author{Gustavo E. Scuseria}
\affiliation{Department of Chemistry, Rice University, Houston, Texas 77005, USA}
\affiliation{Department of Physics and Astronomy, Rice University, Houston, Texas 77005, USA}

\date{\today}

\begin{abstract}
Projected Hartree--Fock theory (PHF) has a long history in quantum chemistry.  PHF is here understood as the variational determination of an $N$-electron broken symmetry Slater determinant that minimizes the energy of a projected state with the correct quantum numbers.  The method was actively pursued for several decades but seems to have been abandoned.  We here derive and implement a ``variation after projection'' PHF theory using techniques different from those previously employed in quantum chemistry.  Our PHF methodology has modest mean-field computational cost, yields relatively simple expressions, can be applied to both collinear and non-collinear spin cases, and can be used in conjunction with  deliberate symmetry breaking and restoration of other molecular symmetries like complex conjugation and point group.  We present several benchmark applications to dissociation curves and singlet-triplet energy splittings, showing that the resulting PHF wavefunctions are of high quality multireference character.  We also provide numerical evidence that in the thermodynamic limit, the energy in PHF is not lower than that of broken-symmetry HF, a simple consequence of the lack of size consistency and extensivity of PHF.
\end{abstract}

\maketitle

\section{Introduction}
In a recent paper,\cite{Scuseria11} we introduced a novel wavefunction method known as projected quasiparticle theory (PQT).  The fundamental idea of PQT is very simple.  We begin with a Hartree--Fock--Bogoliubov determinant (thereby mixing states of differerent particle number) and then restore particle number symmetry with the aid of projection operators in a self-consistent variation-after-projection (VAP) manner.  Without breaking any other symmetries, this reduces to the standard antisymmetrized geminal power (AGP) wavefunction.  However, we can also deliberately break and restore spin symmetry, point group symmetry, and complex conjugation symmetry, all within the same general framework.  In each case, this deliberate symmetry breaking and restoration allows us to construct fairly sophisticated multiconfigurational wavefunctions with mean-field computational scaling.

Closely related to PQT, at least conceptually, is projected Hartree--Fock (PHF), where particle number symmetry is not broken and the projection operators act on an $N$-electron determinant.  The PHF acronym in quantum chemistry is usually associated with spin-projection in a projection-after-variation (PAV) approach\cite{Schlegel86} where the deformed (\textit{i.e.}, symmetry broken) determinant $|\Phi\rangle$ variationally minimizes the energy
\begin{equation}
E_\Phi = \frac{\langle \Phi | \hat{H} | \Phi \rangle}{\langle \Phi|\Phi \rangle},
\end{equation}
and is then used to construct a projected state $|\Psi\rangle = \hat{P} |\Phi\rangle$ whose energy is given by
\begin{equation}
E_\Psi = \frac{\langle \Phi | \hat{P}^\dagger \hat{H} \hat{P} | \Phi \rangle}{ \langle \Phi | \hat{P}^\dagger \hat{P} | \Phi \rangle}
      = \frac{\langle \Phi | \hat{H} \hat{P} | \Phi \rangle}{ \langle \Phi |\hat{P} | \Phi \rangle}.
\label{Eqn:PHF}
\end{equation}
Here, we have used the fact that the projection operator $\hat{P}$ is Hermitian, idempotent, and commutes with the Hamiltonian.  The PAV approach is computationally quite simple, but has several drawbacks.  Near regions of spontaneous symmetry breaking, the PAV energy can be ill behaved.\cite{Mayer78}  Moreover, the projection is often carried out only approximately,\cite{Koga91} \textit{i.e.} without removing all spin contaminants.  This may lead to a large deviation in the expectation value $\langle \hat{S}^2 \rangle$ for the projected state when the reference determinant has contaminants of many different spins.\cite{Tsuchimochi11}  Lastly, the wavefunction is not variationally optimized, complicating the evaluation of properties depending on its derivatives.

These defects can be remedied by using a self-consistent VAP approach, wherein one obtains the deformed determinant $|\Phi\rangle$ by variationally minimizing the energy $E_\Psi$ of Eqn. \ref{Eqn:PHF} with respect to $|\Phi\rangle$, as first proposed by L\"owdin in his extended Hartree--Fock method (EHF).\cite{Lowdin55c}  More often than not, EHF has been associated with the use of a spin projection operator on a reference unrestricted determinant (the so-called spin-projected EHF\cite{Mayer80}), and we will follow this usage. We should note that Goddard's GF method\cite{Goddard68} is equivalent to EHF, while in certain limits also the spin-coupled valence bond (SCVB) method\cite{Byrman93} corresponds to EHF.  The use of projected wavefunctions in quantum chemistry is nowadays very limited, but they are ubiquitous in fields such as nuclear physics.\cite{Ring80,Blaizot85,Schmid2004}

Several authors have proposed different ways to optimize the EHF wavefunction in quantum chemistry.  Mayer's derivation was based on the Brillouin theorem for the projected state.\cite{Mayer73}  His derivation relied heavily on the pairing theorem by L\"owdin.\cite{Mayer10}  Rosenberg and Martino\cite{Rosenberg75}, and later Klimo and Ti\u{n}o\cite{Klimo78}, used a direct minimization of the energy functional. More recently, Byrman suggested an optimization technique based on the fact that the EHF wavefunction is recovered when a single spin wavefunction is used in the context of the spin-coupled valence bond method.\cite{Byrman93}  Our purpose in this paper is to show how to carry out a VAP optimization of a PHF wavefunction in an alternative, computationally efficient, manner.  In achieving this goal, we are guided by our previous work.\cite{Scuseria11}

The derivation of the PQT equations relied on expressing a general Hartree--Fock--Bogoliubov determinant $|\Phi \rangle$ using its Thouless parametrization with respect to the bare vacuum $|-\rangle$.  In other words, we implicitly wrote $|\Phi \rangle = e^{\hat{Z}} |-\rangle$.\cite{Scuseria11,Sheikh2000}  This parametrization requires $\langle \Phi | - \rangle \neq 0$.  In the case of PHF, however, the deformed determinant is a number eigenfunction with $N$ electrons and is thus orthogonal to the bare vacuum.  The PQT derivation cannot then be blindly applied.  The practical consequence of this orthogonality is that one encounters indeterminancies in the PQT equations when using reference determinants whose one-particle density matrix has occupation numbers equal to zero.

While the PQT derivation cannot straightforwardly be extended to derive corresponding PHF equations, the underlying conceptual framework is similar.  That is, we wish to express the energy of the projected state in terms of the one-particle density matrix of a broken symmetry reference determinant.  Having done so, the energy can be minimized with respect to the reference determinant's density matrix, leading to an effective one-particle Hamiltonian to be variationally optimized.  We thus follow the basic ideas used in PQT to derive the PHF equations from scratch.  Our derivation and resulting equations afford spin symmetry restoration of both collinear and non-collinear broken symmetry determinants in a straighforward manner, where collinear (unrestricted Hartree--Fock-type) determinants are eigenfunctions of $\hat{S}_z$ and non-collinear determinants (generalized Hartree--Fock-type) are not.  Additionally, we can readily restore complex conjugation or point group symmetry as well, just as we did in PQT.  All of this is accomplished in essence by choosing a more convenient representation of the projection than the standard L\"owdin projection operators used historically.

The use of non-collinear reference determinants in the optimization of projected Hartree--Fock states has been limited to the work by Lunell on the two-electron series.\cite{Lunell70} The complex molecular orbital method of Hendekovi\'c\cite{Hendekovic74} is closely related to our complex conjugation projection but has seen little use.   Unlike these approaches, however, our implementation of PHF readily allows for the simultaneous restoration of many symmetries at once.  Our method is also related to the VAMPIR method of Schmid \textit{et al.},\cite{Schmid2005} though the optimization of the reference determinant is carried out in a very different way.

The properties of the projected Hartree--Fock wavefunction have been studied extensively by many authors.  We point the interested reader to the review by Mayer\cite{Mayer80} on the spin-projected EHF method as a starting point to access the wide literature on the subject.  In particular, we mention that properties of the density matrices characterizing the standard EHF states (\textit{i.e.}, those based on collinear determinants) have been derived by Harriman,\cite{Harriman64} Hardisson and Harriman,\cite{Hardisson67}, Mestechkin,\cite{Mestechkin73} and Phillips and Schug.\cite{Phillips74}  Simons and Harriman\cite{Simons69} derived properties of the point-group projected Hartree--Fock wavefunction.

We should also point out some of the vices of PHF: the method is neither size consistent nor size extensive, as it has been noted before.\cite{Castano86}  Recall that a size consistent method is one where the dissociation limit of the system $AB$ is equal to the energy of $A$ plus the energy of $B$; a size-extensive method is one where the energy is proportional to the number of particles.  Our results in this paper confirm previous conclusions\cite{Mayer76,Castano86} that, in the thermodynamic limit, the PHF energy per particle reduces to that of the broken symmetry mean field. In other words, as the number of electrons becomes large enough, PHF has nothing to add over standard Hartree--Fock except that it retains good quantum numbers which the broken symmetry Hartree--Fock state loses.

Section \ref{Sec:Theory} discusses the details of the projection we use before deriving the energy of our projected wavefunction and the variational equations we solve to obtain it.  In Sec. \ref{Sec:Results} we present the results of applying our equations in the description of molecular dissociation processes and in the computation of singlet-triplet splittings, before drawing a few conclusions in Sec. \ref{Sec:Conclusions}.  In the interest of brevity, we have deferred the detailed equations for our effective Hamiltonian to the appendix.

\section{Theory
\label{Sec:Theory}}

\subsection{Projection Operators}
The complicated nature of Mayer's EHF equations owes much to the form of the projection operator used.  This form is due to L\"owdin,\cite{Lowdin55c} and writes
\begin{equation}
\hat{P}^s = \prod_{l \neq s} \frac{\hat{S}^2 - l(l+1)}{s(s+1) -  l(l+1)}.
\end{equation}
Applying this operator to a wavefunction composed of multiple different spins just returns the component with spin $s$, and annihilates the components with other spins.  While this spin operator does not project the $z$-component of spin, it can be generalized to do so.  Given a simpler way of writing the projection operators and evaluating the projected energy $E_\Psi$ in terms of the deformed determinant $|\Phi\rangle$, one would have a computationally simpler scheme whose results are identical to those of Mayer in the case of spin projection.

Fortunately, such a simple representation of general projections exists.\cite{Ring80,Blaizot85}  In our previous paper,\cite{Scuseria11} we omitted some details about the form of the projection, a fact we intend to remedy here. We start our discussion by emphasizing that by projected Hartree--Fock theory (PHF) we mean recovering a wavefunction with good quantum numbers from an intrinsically deformed HF state, and do not demand the use of actual projection operators in a strict mathematical sense.  In other words, while we require that the wavefunction $\hat{P} | \Phi\rangle$ be an eigenfunction of the relevant symmetry operators, we do not insist that $\hat{P}$ is either Hermitian or idempotent.

If a single operator $\hat{\Lambda}$ can be associated with a constant of motion (that is, $\hat{\Lambda}$ commutes with the Hamiltonian), one can write a projection operator as
\begin{equation}
\hat{P}^\lambda = \frac{1}{L} \int_L \mathrm{d}\phi \,  e^{\mathrm{i} \phi (\hat{\Lambda} - \lambda)},
\end{equation}
where $\lambda$ is the eigenvalue recovered and $L$ is the corresponding volume of integration.\cite{Ring80}  We note that this form can be used to project onto eigenfunctions of $\hat{S}_z$.

If, on the other hand, there is a set of operators that commute with the Hamiltonian (but not necessarily with each other) and which form a group $G = \{\hat{g}\}$, then one can recover the correct symmetries by diagonalizing the Hamiltonian in the basis formed by the elements of the group.  In other words,
\begin{equation}
|\Psi \rangle = \sum_g c_g \, \hat{g} |\Phi \rangle,
\end{equation}
with the coefficients being determined by the solution to the corresponding eigenvalue problem.  

Alternatively, one can recover the coefficients by demanding that the wavefunction $|\Psi\rangle$ has the desired symmetries.  In other words, one can work with projectors.  In general, one can construct the so-called ``transfer'' operators\cite{Tinkham,Lowdin66}
\begin{equation}
\hat{P}^j_{\lambda \kappa} = \frac{l_j}{h} \sum_g \Gamma^j (g)^\star_{\lambda \kappa} \, \hat{g},
\end{equation}
where $h$ is the order of the group $G$, $l_j$ is the dimension of the irreducible representation $\Gamma^j$, and $\Gamma^j (g)_{\lambda \kappa}$ is the element in the $\lambda$-th row and $\kappa$-th column of the matrix associated with $\hat{g}$ in such irreducible representation. This operator yields zero unless the function on which it acts belongs to the $\kappa$-th row of $\Gamma^j$.  By using the great orthogonality theorem one can show that
\begin{align}
\hat{P}^j_{\lambda \kappa} \hat{P}^{k}_{\mu \nu} &= \delta_{jk} \delta_{\kappa \mu} \hat{P}^{k}_{\lambda \nu},
 \\
(\hat{P}^j_{\lambda \kappa})^\dagger &= \hat{P}^j_{\kappa \lambda}
\end{align}
It is easy to see that $\hat{P}^j_{\kappa \kappa}$ is indeed a projection operator in the mathematical sense: it is Hermitian and idempotent.

Consider the action of $\hat{P}^j_{\kappa \kappa}$ on a deformed wavefunction $|\Phi \rangle$.  It extracts the component of $|\Phi \rangle$ which transforms as the $\kappa$-th row of $\Gamma^j$.  It is, however, unphysical in the sense that
\begin{equation}
\hat{P}^j_{\kappa \kappa} |\Phi \rangle \neq \hat{P}^j_{\kappa \kappa} |\Phi' \rangle
\end{equation}
where $|\Phi' \rangle$ is a rotated wavefunction in the subspace of $\Gamma^j$.  In order to avoid this unphysical behavior, we use the linear combination
\begin{equation}
|\Psi \rangle = \sum_\kappa c_\kappa \hat{P}^j_{\lambda \kappa} |\Phi \rangle.
\end{equation}
It is clear that this linear combination produces a wavefunction $|\Psi \rangle$ which transforms as the $\lambda$-th row of $\Gamma^j$, thus having $\lambda$ and $j$ as good quantum numbers.

Evaluating the energy of $|\Psi \rangle$ leads to
\begin{align}
E_\Psi
  &= \frac{\sum_{\kappa \kappa'} c_{\kappa'}^\star \, c_\kappa \langle \Phi | (\hat{P}^j_{\lambda \kappa'})^\dagger \hat{H} \hat{P}^j_{\lambda \kappa} | \Phi \rangle}
          {\sum_{\kappa \kappa'} c_{\kappa'}^\star \, c_\kappa \langle \Phi | (\hat{P}^j_{\lambda \kappa'})^\dagger \hat{P}^j_{\lambda \kappa} | \Phi \rangle}
\nonumber
\\
  &= \frac{\sum_{\kappa \kappa'} c_{\kappa'}^\star \, c_\kappa \langle \Phi | \hat{H} \hat{P}^j_{\kappa' \kappa} | \Phi \rangle}
          {\sum_{\kappa \kappa'} c_{\kappa'}^\star \, c_\kappa \langle \Phi | \hat{P}^j_{\kappa' \kappa} | \Phi \rangle},
\end{align}
where we have used the fact that $\hat{P}^j_{\lambda \kappa}$ commutes with the Hamiltonian.  The coefficients $c_\kappa$ are most conveniently determined by solving the generalized eigenvalue problem
\begin{equation}
\sum_{\kappa} h^j_{\kappa' \kappa} \, c_{\kappa} = E \sum_{\kappa} n^j_{\kappa' \kappa} \, c_{\kappa},
\end{equation}
with $h^j_{\kappa' \kappa} = \langle \Phi | \hat{H} \hat{P}^j_{\kappa' \kappa} | \Phi \rangle$ and $n^j_{\kappa' \kappa} = \langle \Phi | \hat{P}^j_{\kappa' \kappa} | \Phi \rangle$.  Observe that all projected wavefunctions of the same irreducible representation $\Gamma^j$ are degenerate.

We have used the above projections to restore spatial symmetry.  On the other hand, we have preferred to diagonalize the Hamiltonian in the basis of the elements of the group $\{ \hat{I},\hat{K} \}$ to restore complex conjugation.  Restoring complex conjugation amounts to letting the resulting $| \Psi \rangle$ have the property that
\begin{equation}
\hat{K} |\Psi \rangle = e^{\mathrm{i} \chi} |\Psi \rangle,
\end{equation}
where $\chi$ is an arbitrary phase factor.  Note that a general complex deformed HF determinant does not have this property.

\subsection{Spin Projection Operators}
Due to the historical significance of spin projection, we here present a more detailed discussion of the form we use.

We can recover eigenfunctions of $\hat{S}^2$  by demanding that the projected wavefunction be invariant to the axis of spin quantization, as first proposed by Percus and Rotenberg in the context of angular momentum projection.\cite{Percus62}  The form we use for the spin projection operator is similar to the one we discussed for non-abelian groups in the preceding section. Explicitly, we use the operator
\begin{align}
\hat{P}^s_{mk} 
  &= | s;m \rangle \langle s;k|
\\
  &= \frac{2s + 1}{8\pi^2} \int \mathrm{d}\Omega \,  D^{s\star}_{mk} (\Omega) \, \hat{R} (\Omega),
\\
\hat{R}(\Omega)
  &= \mathrm{e}^{\mathrm{i} \alpha \hat{S}_z} \mathrm{e}^{\mathrm{i} \beta \hat{S}_y}  \mathrm{e}^{\mathrm{i} \gamma \hat{S}_z},
\end{align}
where $D^s_{mk}(\Omega) = \langle s;m| \hat{R}(\Omega) | s;k\rangle$ is a Wigner rotation matrix and $|s;m\rangle$ is an eigenfunction of $\hat{S}^2$ and $\hat{S}_z$ with eigenvalues denoted by $s$ and $m$, respectively.  Here, $\Omega = (\alpha, \beta, \gamma)$ is the set of Euler angles.  

$\hat{P}^s_{mk}$ obeys the same rules as the transfer operators discussed previously:
\begin{align}
\hat{P}^s_{mk} \, \hat{P}^{s'}_{m'k'} &= \delta_{ss'} \, \delta_{m'k} \, \hat{P}^s_{mk'},
\\
\left( \hat{P}^s_{mk} \right)^\dagger &= \hat{P}^s_{km}.
\end{align}
The action of $\hat{P}^s_{mk}$ on a deformed HF state $|\Phi \rangle$ recovers the multi-determinantal state characterized by the quantum numbers $s$ and $k$ and writes it in terms of the quantum numbers $s$ and $m$.  As discussed above, using such an operator is undesirable as the wavefunction (and the energy) obtained depends on the value of $k$ chosen.  We therefore form the linear combination
\begin{equation}
|\Psi \rangle = \sum_k c_k \hat{P}^s_{mk} |\Phi \rangle
\end{equation}
and determine the $2s+1$ coefficients $c_k$ by diagonalization.

We finally note that if $|\Phi \rangle$ is an eigenfunction of $\hat{S}_z$ with eigenvalue $m$ (\textit{i.e.}, a collinear state), then no such expansion is necessary.  Furthermore, the operator becomes a true projection operator and can be simplified to
\begin{equation}
\hat{P}^s_{mm} = \frac{2s+1}{2} \int_0^\pi \mathrm{d}\beta \, \sin \beta \, d^s_{mm} (\beta) \, e^{i\beta \hat{S}_y},
\end{equation}
where $d^s_{mm} (\beta) = \langle s;m| e^{i\beta \hat{S}_y} | s;m\rangle$.  In this case, the projected wavefunctions (and their energies) depend on the value of $\langle \hat{S}_z \rangle = m$ chosen for the deformed determinant.  In practice, the gauge integration over the set of Euler angles is discretized.  We have observed that the number of grid points required to get good convergence on the integrals is generally small.  We should note that the projectors we have described in the foregoing have frequently been used for total angular momentum projection in nuclear physics.\cite{Ring80}  It should also be observed that the form of the spin projector that we use is in fact not new in quantum chemistry.  It was used by Lefebvre and Prat\cite{Lefebvre67,Lefebvre69} to provide a simpler formula for the EHF energy.  Their work, unfortunately, remained largely unnoticed.

To reduce notational clutter, we will henceforth use $E$ for the energy of the projected state and will write the projector in a general form as
\begin{equation}
\hat{P} = \int \mathrm{d} \Omega \, w(\Omega) \, \hat{R}(\Omega).
\end{equation}
All associated eigenvalue problems are implied by our notation.

\subsection{Projected Hartree--Fock}
Given the projectors defined above, we evaluate the projected Hartree--Fock energy expression as
\begin{equation}
E = \frac{\langle \Phi | \hat{H} \hat{P} | \Phi \rangle}
         {\langle \Phi | \hat{P} | \Phi \rangle}
  = \frac{\int\mathrm{d}\Omega \,  w(\Omega) \, \langle \Phi | \hat{H} \hat{R}(\Omega) | \Phi\rangle}
         {\int\mathrm{d}\Omega \, w(\Omega) \, \langle \Phi | \hat{R}(\Omega)  | \Phi\rangle}.
\end{equation}
Simplifying the notation by defining
\begin{equation}
x(\Omega) = w(\Omega) \, \langle \Phi | \hat{R}(\Omega) | \Phi\rangle
\end{equation}
and constructing the rotated Slater determinants
\begin{equation}
|\Omega\rangle = \frac{\hat{R}(\Omega) | \Phi\rangle}{\langle \Phi | \hat{R}(\Omega) | \Phi \rangle}
\end{equation}
permits us to write
\begin{equation}
E = \frac{\int\mathrm{d}\Omega \, x(\Omega) \, \langle 0 | \hat{H} | \Omega \rangle}
         {\int\mathrm{d}\Omega \, x(\Omega)} 
  = \int\mathrm{d}\Omega \, y(\Omega) \, \langle 0 | \hat{H} | \Omega \rangle.
\label{Eqn:EPHF}
\end{equation}
Evaluating the energy thus requires evaluating overlap or norm matrix elements $\langle \Phi | \hat{R}(\Omega) | \Phi\rangle$ and Hamiltonian matrix elements $\langle 0 | \hat{H} | \Omega \rangle$.  Note that because $\hat{R}(0) = 1$, we have $|0\rangle = |\Phi\rangle$.  Note also that the determinants $|\Omega\rangle$ are defined in intermediate normalization, so that $\langle 0 | \Omega\rangle = 1$.

The norm matrix elements can be evaluated in the usual way for the overlap between two non-orthogonal Slater determinants $|\Phi\rangle$ and $|\Xi\rangle$:\cite{Lowdin55b}
\begin{equation}
\langle \Phi | \Xi \rangle = \mathrm{det}\,\mathbf{M},
\end{equation}
where $\mathbf{M}$ is the matrix of overlaps between orbitals $|\phi\rangle$ occupied in $|\Phi\rangle$ and orbitals $|\xi\rangle$ occupied in $|\Xi\rangle$, \textit{i.e.}
\begin{equation}
M_{ij} = \langle \phi_i | \xi_j\rangle.
\end{equation}
To evaluate the Hamiltonian matrix elements we follow L\"owdin,\cite{Lowdin55b} who realized that a generalized form of Wick's theorem holds.  We have
\begin{equation}
\frac{\langle \Phi | \hat{H} |\Xi\rangle}{\langle \Phi | \Xi \rangle}
   = \sum h_{ik} \rho_{ki}^{\Phi\Xi} + \frac{1}{2} \sum \langle ij \| kl \rangle \rho_{ki}^{\Phi\Xi} \rho_{lj}^{\Phi\Xi}
\end{equation}
where $h_{ik}$ and $\langle ij || kl \rangle$ are the usual one-electron and antisymmetrized two-electron integrals, respectively, and the transition density matrix elements $\rho_{kl}^{\Phi\Xi}$ are given by
\begin{equation}
\rho_{kl}^{\Phi\Xi} = \frac{\langle \Phi| a_l^\dagger a_k |\Xi\rangle}{\langle\Phi|\Xi\rangle}
                 = \sum_{i,j=1}^N \langle k|\xi_i\rangle (\mathbf{M}^{-1})_{ij} \langle \phi_j | l\rangle.
\end{equation}

Using these formulae, we find that the overlap we need is given by
\begin{align}
\langle\Phi| \hat{R}(\Omega) | \Phi\rangle
  &= \mathrm{det}\,\mathbf{M}_\Omega
\\
\mathbf{M}_\Omega
  &= \mathbf{C}^\dagger \mathbf{R}_\Omega \mathbf{C},
\label{Eqn:DefM}
\end{align}
while the Hamiltonian elements are
\begin{equation}
\langle 0 | \hat{H} | \Omega \rangle 
= \sum h_{ik} (\bm{\rho}_\Omega)_{ki} + \frac{1}{2} \sum \langle ij \| kl\rangle (\bm{\rho}_\Omega)_{ki} (\bm{\rho}_\Omega)_{lj}.
\end{equation}
The transition density matrices
\begin{equation}
(\bm{\rho}_\Omega)_{kl} = \frac{\langle \Phi | a_l^\dagger a_k  \hat{R}(\Omega) | \Phi\rangle}{\langle \Phi| \hat{R}(\Omega) | \Phi\rangle}
\end{equation}
can be formed as
\begin{equation}
\bm{\rho}_\Omega = \mathbf{R}_\Omega \mathbf{C} \mathbf{M}^{-1}_\Omega \mathbf{C}^\dagger.
\label{Eqn:DefRho}
\end{equation}
Here and above, $\mathbf{C}$ is the $M \times N$ matrix of orbital coefficients defining the occupied orbitals in $|\Phi\rangle$ and $\mathbf{R}_\Omega$ is the matrix representation of the operator $\hat{R}(\Omega)$, where $M$ is the number of spin-orbitals in the basis and $N$ is the number of electrons in the system.  We have assumed an orthonormal basis for convenience and will continue to do so, but modifications for a non-orthonormal basis are straightforward.

\subsection{Optimization of the PHF Wave Function
\label{Sec:Optimization}}
As we have written things thus far, the fundamental variables are the occupied orbital coefficients $\mathbf{C}$.  However, because the wavefunction $|\Phi\rangle$ is a single determinant, it is defined completely by its one-particle density matrix $\bm{\rho}$, a point brought up by L\"owdin in 1966,\cite{Lowdin66} when he wrote
\begin{quote}
``Since the wavefunction $\hat{P} | \Psi\rangle$ depends uniquely on $\bm{\rho}$, the main problem in the projected Hartree--Fock scheme is to vary $\bm{\rho}$ so that the energy becomes an absolute minimum.''
\end{quote}
The minimization of the energy with respect to $\rho$ would be greatly facilitated if we were able to write an explicit density matrix functional $E[\rho]$.  Our task here is thus to obtain this energy functional, in a manner analogous to the work of Sheikh and Ring for projected Hartree--Fock--Bogoliubov.\cite{Sheikh2000}  Having done so, we can minimize the PHF energy with respect to idempotent density matrices $\bm{\rho}$. 

We start by writing the density matrix as
\begin{equation}
\bm{\rho} = \mathbf{C} \mathbf{C}^\dagger
\end{equation}
where we recall that $\mathbf{C}$ is the matrix of coefficients defining the occupied orbitals.  Let us partition this matrix in the form
\begin{equation}
\mathbf{C} = \begin{pmatrix} \mathbf{C}_p \\ \mathbf{C}_q \end{pmatrix}
\end{equation}
where $\mathbf{C}_p$ is $N \times N$ and $\mathbf{C}_q$ is $(M-N) \times N$.  In terms of these blocks of the orbital coefficients, we have
\begin{align}
\bm{\rho}
  &= \begin{pmatrix} \bm{\rho}_{pp} & \bm{\rho}_{pq} \\  \bm{\rho}_{qp} & \bm{\rho}_{qq} \end{pmatrix}
\\
  &= \begin{pmatrix} \mathbf{C}_p \, \mathbf{C}_p^{\dagger}  & \mathbf{C}_p \, \mathbf{C}_q^{\dagger} \\
                     \mathbf{C}_q \, \mathbf{C}_p^{\dagger}  & \mathbf{C}_q \, \mathbf{C}_q^{\dagger}\end{pmatrix}.
\end{align}
Note that the partitioning of $\mathbf{C}$ is not unique, because we can arbitrarily mix basis functions amongst each other, thereby mixing rows of $\mathbf{C}$.  We require that $\mathbf{C}_p$ have non-vanishing determinant.  We can always find a partitioning satisfying this requirement because we can choose to work in the basis of the occupied molecular orbitals of $|\Phi\rangle$ for which $\mathbf{C}_p = \mathbf{1}$.  In practical calculations, we do exactly this, and the subscripts $p$ and $q$ denote occupied and virtual orbitals, respectively.

Recall that the overlap matrix elements are given by
\begin{equation}
\langle \Phi | \hat{R}(\Omega) | \Phi \rangle = \mathrm{det} \, \mathbf{M}_\Omega.
\end{equation}
We rewrite the matrix $\mathbf{M}_\Omega$ as
\begin{subequations}
\begin{align}
\mathbf{M}_\Omega
  &= \begin{pmatrix} \mathbf{C}_p \\ \mathbf{C}_q \end{pmatrix}^\dagger \mathbf{R}_\Omega 
     \begin{pmatrix} \mathbf{C}_p \\ \mathbf{C}_q \end{pmatrix}
\\
  &= (\mathbf{C}_p)^{-1} \mathbf{C}_p 
      \begin{pmatrix} \mathbf{C}_p \\ \mathbf{C}_q \end{pmatrix}^\dagger \mathbf{R}_\Omega 
      \begin{pmatrix} \mathbf{C}_p \\ \mathbf{C}_q \end{pmatrix}
      \mathbf{C}_p^\dagger (\mathbf{C}_p^\dagger)^{-1}
\\
  &= (\mathbf{C}_p)^{-1} \mathbf{N}_\Omega^{-1} (\mathbf{C}^\dagger_p)^{-1},
\end{align}
\end{subequations}
where
\begin{equation}
\mathbf{N}_\Omega = 
\left[ \begin{pmatrix} \bm{\rho}_{pp} & \bm{\rho}_{pq} \end{pmatrix} \mathbf{R}_\Omega  
       \begin{pmatrix} \bm{\rho}_{pp} \\ \bm{\rho}_{qp}\end{pmatrix} \right]^{-1}.
\label{defN}
\end{equation}
Then we can write the overlap matrix elements explicitly in terms of $\bm{\rho}$ as
\begin{equation}
\langle \Phi | \hat{R}(\Omega) | \Phi \rangle = \frac{1}{\mathrm{det}(\mathbf{N}_\Omega \, \bm{\rho}_{pp})}.
\end{equation}

In a similar manner, we can write the Hamiltonian matrix elements, computed in terms of transition density matrices, explicitly in terms of $\bm{\rho}$. The transition density matrices are formed as
\begin{equation}
\bm{\rho}_\Omega = \mathbf{R}_\Omega     \begin{pmatrix} \mathbf{C}_p \\ \mathbf{C}_q \end{pmatrix}
                 \mathbf{M}^{-1}_\Omega \begin{pmatrix} \mathbf{C}_p \\ \mathbf{C}_q \end{pmatrix}^\dagger.
\end{equation}
Simple manipulations bring us to
\begin{equation}
\bm{\rho}_\Omega = \mathbf{R}_\Omega \begin{pmatrix} \bm{\rho}_{pp}  \\ \bm{\rho}_{qp} \end{pmatrix} \mathbf{N}_\Omega 
                                  \begin{pmatrix} \bm{\rho}_{pp} & \bm{\rho}_{pq} \end{pmatrix}.
\label{Def:TransitionDensityMatrices}
\end{equation}

Having expressed the overlap and Hamiltonian matrix elements as functionals of $\bm{\rho}$, we have all the ingredients we need to write a closed-form expression for the projected energy in terms of $\bm{\rho}$, which we recall is the density matrix of the deformed (unprojected) state $|\Phi\rangle$.  We can then set up a variational problem similar to that of the regular Hartree--Fock procedure.  We simply write
\begin{equation}
\delta \{ E[\bm{\rho}] - \mathrm{Tr}[\bm{\Lambda} (\bm{\rho}^2 -  \bm{\rho})]\} = 0,
\end{equation}
where $\bm{\Lambda}$ is a matrix of Lagrange multipliers used to constrain $\bm{\rho}$ to remain idempotent so that $|\Phi\rangle$ remains a single determinant.  This variational ansatz is equivalent to the condition that
\begin{equation}
[\bm{\mathcal{F}},\bm{\rho}] = 0,
\label{Brillouin}
\end{equation}
where $\bm{\mathcal{F}}$ is an effective Fock matrix given by
\begin{equation}
\mathcal{F}_{kl} = \frac{\partial\hfill}{\partial \rho_{lk}} E[\bm{\rho}].
\end{equation}
We provide explicit expressions for the matrix elements of $\bm{\mathcal{F}}$ in the appendix.

Equation $\ref{Brillouin}$ can be solved by what we may call the PHF eigenvalue equations
\begin{equation}
\bm{\mathcal{F}} \mathbf{C} = \mathbf{C} \bm{\epsilon}
\end{equation}
where $\bm{\epsilon}$ is a (diagonal) matrix of orbital energies and $\mathbf{C}$ here corresponds to the full matrix of orbital coefficients (including virtual orbitals).  We use the eigenvectors $\mathbf{C}$ of $\bm{\mathcal{F}}$ to construct a new guess of the density matrix $\bm{\rho}$, just as in standard Hartree--Fock.  We remind the reader that these equations have been here derived in the special case of an orthonormal basis and must be modified slightly in the case of a general non-orthogonal basis.

Thus, the optimization of the projected Hartree--Fock state uses the following algorithm:
\begin{enumerate}
  \item Construct an initial broken symmetry guess for $|\Phi\rangle$ and form $\bm{\rho}$.
  \item Compute the overlap matrix elements $\langle \Phi | \hat{R}(\Omega) | \Phi \rangle$ and form the function $y(\Omega)$.
  \item Form the transition density matrices $\bm{\rho}_\Omega$ and contract them with one- and two-electron integrals to evaluate the Hamiltonian matrix elements $\langle 0 | \hat{H} | \Omega \rangle$.
  \item Form the effective Fock matrix $\bm{\mathcal{F}}$ and diagonalize it to obtain new eigenvectors and thence a new density matrix $\bm{\rho}$.
  \item Test for convergence.  If the density matrix is not converged, return to step 2.
\end{enumerate}
At any time the energy can be evaluated from
\begin{equation}
E = \int\mathrm{d}\Omega \, y(\Omega) \, \langle 0 | \hat{H} | \Omega \rangle.
\end{equation}

We emphasize that the scaling of our PHF implementation with respect to system size is the same as that of Hartree--Fock.  At each gauge point $\Omega$, we form an effective one-body Hamiltonian $\bm{\mathcal{F}}_\Omega$ by contracting two-electron integrals with transition density matrices (whose formation is trivial).  Having formed $\bm{\mathcal{F}}$ by integrating $\bm{\mathcal{F}}_\Omega$ over the gauge angle $\Omega$, we obtain a new reference determinant by diagonalizing $\bm{\mathcal{F}}$.  While the cost of our PHF procedure is moderately higher than that of Hartree--Fock because we must integrate over the gauge angle, this integration generally does not require too many points and is trivially parallelizable in any event.  Reference \onlinecite{Scuseria11} establishes that the size of the gauge integration grid needed scales only weakly with the size of the system.

It is important to note that the Brillouin-like condition of Eqn. \ref{Brillouin} defines only the occupied-virtual part of the effective Hamiltonian $\bm{\mathcal{F}}$, just as in standard Hartree--Fock theory.  When working in the molecular orbital basis as we do, this corresponds to defining $\bm{\mathcal{F}}_{pq}$ and $\bm{\mathcal{F}}_{qp}$.  The occupied-occupied and virtual-virtual parts of $\bm{\mathcal{F}}$ are not defined by the Brillouin condition; accordingly, without further modifications the entire effective Hamiltonian vanishes at convergence.  In order to cleanly separate occupied and virtual orbitals and thereby ensure efficient convergence, the occupied-occupied and virtual-virtual parts must be defined as well.  We therefore define the occupied-occupied and virtual-virtual parts of $\bm{\mathcal{F}}$ to be the corresponding parts of $\mathbf{F}$, the standard Fock operator constructed from the density matrix $\bm{\rho}$.  In the molecular orbital basis, this corresponds to defining $\bm{\mathcal{F}}_{pp} = \mathbf{F}_{pp}$ and $\bm{\mathcal{F}}_{qq} = \mathbf{F}_{qq}$; in a general basis, this corresponds to writing $\bm{\mathcal{F}} \to \bm{\mathcal{F}} + \bm{\rho} \mathbf{F}\bm{\rho} + (\bm{1}-\bm{\rho}) \mathbf{F} (\bm{1} - \bm{\rho})$.  This \textit{ad hoc} choice has proven to be efficient and reliable thus far but is not guaranteed to work in all cases.  One should note that the orbital energies from our PHF equations do not have obvious physical meaning.

\subsection{Nomenclature}
Before we discuss our results, let us briefly clarify the nomenclature we use in this manuscript.

We will use PUHF for spin projection on the unrestricted Hartree--Fock (UHF) in a projection after variation manner.  We write SUHF (equivalent to the standard spin-projected EHF) for variation after spin projection on a determinant which breaks $\hat{S}^2$ symmetry and SGHF for variation after spin projection on a determinant which breaks both $\hat{S}^2$ and $\hat{S}_z$ symmetry.  We can also break complex conjugation symmetry (see Ref. \onlinecite{Scuseria11}) to form KUHF, KGHF, KSUHF, or KSGHF, where KSGHF, for example, means that we restore complex conjugation symmetry as well as $\hat{S}^2$ and $\hat{S}_z$.  Similarly, we can restore point group symmetry to form, for instance, C$_2$-SUHF, which restores $\hat{S}^2$ and makes the wavefunction transform as one of the irreducible representations of the group $C_2$.

\section{Results
\label{Sec:Results}}
Our PHF equations have been implemented in the \textsc{Gaussian}\cite{Gaussian} program package.  For this pilot work, we will generally be interested in spin projection in small systems with small basis sets, though we also consider complex conjugation and point group symmetry breaking and restoration.  Larger systems with larger basis sets are certainly feasible, as the method has the same computational scaling as Hartree--Fock.

In order to construct an initial guess for the deformed determinant, we typically start with a broken-symmetry Hartree--Fock state.  This is usually straightforward in the case of SUHF, but is often challenging for SGHF since GHF solutions different from rotated UHF are not common for first row molecules.\cite{Jimenez11}  We therefore mix the $\uparrow$-spin and $\downarrow$-spin orbitals closest to the Fermi level with some predefined angle small enough so as not to raise the energy significantly but large enough so as not to return to UHF.  Similarly, to break complex conjugation symmetry, we mix the highest occupied orbital and the lowest virtual orbital with complex coefficients, thereby breaking complex conjugation symmetry, and we use an analogous procedure to break point group symmetry where we must take care to mix orbitals belonging to different irreducible representations of the point group we wish to restore.  Convergence is usually enhanced by the use of the Direct Inversion of the Iterative Subspace (DIIS)\cite{Pulay82} procedure which ideally does not start until the energy of the projected state is below the energy of the symmetry-adapted HF state.  In other words, SUHF projecting onto an overall singlet state should not start DIIS until the energy is lower than the RHF energy, simply because the RHF wavefunction is a solution of the SUHF equations.

Our first order of business is to verify that our SUHF method reproduces previous EHF results.  Unfortunately, there are relatively few calculations available in the literature.  Nonetheless, we have verified that we reproduce results from Rosenberg and Martino\cite{Rosenberg75} and Klimo and Ti\u{n}o\cite{Klimo78}. Recently, Karadakov and Cooper performed self-consistent projected UHF calculations based on the spin-coupled valence bond theory.\cite{Karadakov09}  Because they worked within the SCVB framework, Karadakov and Cooper chose to keep a set of core orbitals which remained doubly occupied and symmetry adapted. The wavefunction is then the antisymmetrized product of the core part with an active part, written as the product of a set of non-orthogonal spatial orbitals with a single spin function, whose coefficients are calculated variationally.  In order to make our results directly comparable to theirs, we had to carry out a constrained optimization of our SUHF state in a manner similar to our previous work on CUHF, such that symmetry breaking is only allowed within an active space.\cite{Tsuchimochi10,Tsuchimochi11}  We have reproduced several of the numbers quoted in Ref. \onlinecite{Karadakov09} to all decimal places.

\subsection{Molecular Dissociation}
We start by examining the effects of projection on molecular dissociation curves.  All calculations in this section use the cc-pVDZ basis set,\cite{Dunning89} unless otherwise stated.

\begin{figure}
\includegraphics[width=\columnwidth]{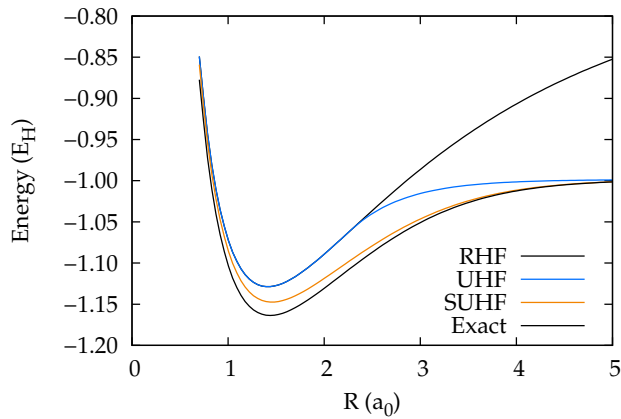}
\caption{Potential energy curve for H$_2$ dissociation in the cc-pVDZ basis set.
\label{Fig:H2Diss}}
\end{figure}

As is well known, restricted Hartree--Fock (RHF) fails for the dissociation of H$_2$, separating to an equal mix of two hydrogen atoms on the one hand and a hydrogen cation plus a hydrogen anion on the other.  The energetically correct dissociation limit is recovered by breaking spin and spatial symmetry in UHF.  The resulting wavefunction is a linear combination of singlet and triplet and is thus spin contaminated.  Restoring spin symmetry variationally leads to the SUHF wavefunction, which like UHF is energetically exact at dissociation but unlike UHF retains good quantum numbers along the whole potential energy curve.  Near equilibrium, SUHF and UHF differ significantly, as can be seen in Fig. \ref{Fig:H2Diss}.  Breaking and restoring symmetry under $\hat{K}$ or $\hat{S}_z$ variationally improves the results near equilibrium even further.

\begin{figure}
\includegraphics[width=\columnwidth]{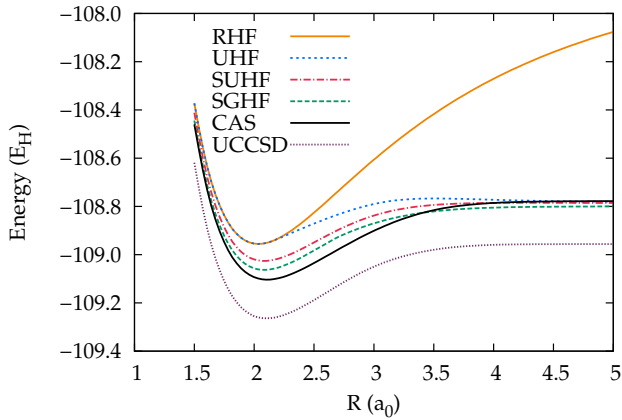}
\caption{Effects of spin projection on the potential energy curves for N$_2$ dissociation in the cc-pVDZ basis set, compared to the CASSCF(10/8) reference.
\label{Fig:N2Diss1}}
\end{figure}

\begin{figure}
\includegraphics[width=\columnwidth]{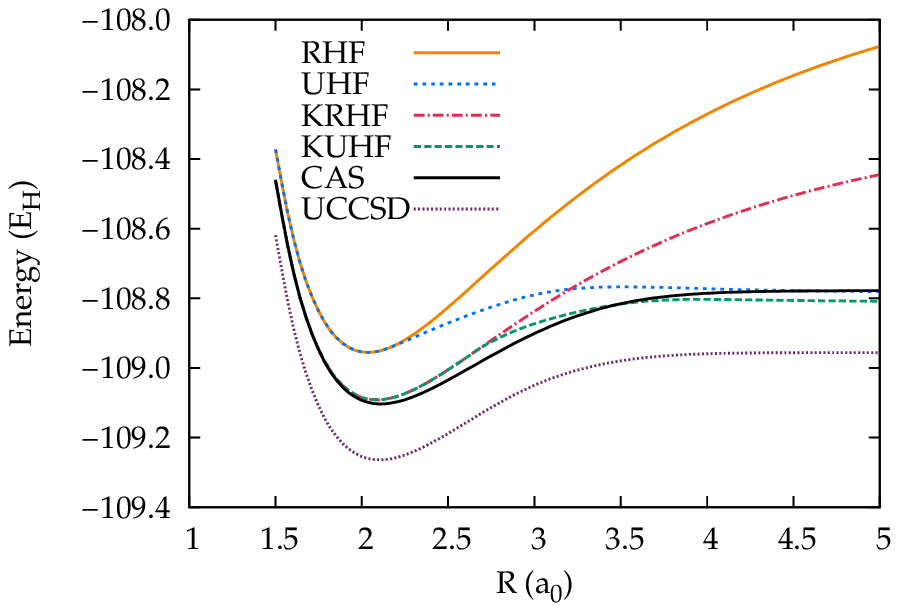}
\caption{Effects of complex conjugation projection on the potential energy curves for N$_2$ dissociation in the cc-pVDZ basis set, compared to the CASSCF(10/8) reference.
\label{Fig:N2Diss2}}
\end{figure}

\begin{figure}
\includegraphics[width=\columnwidth]{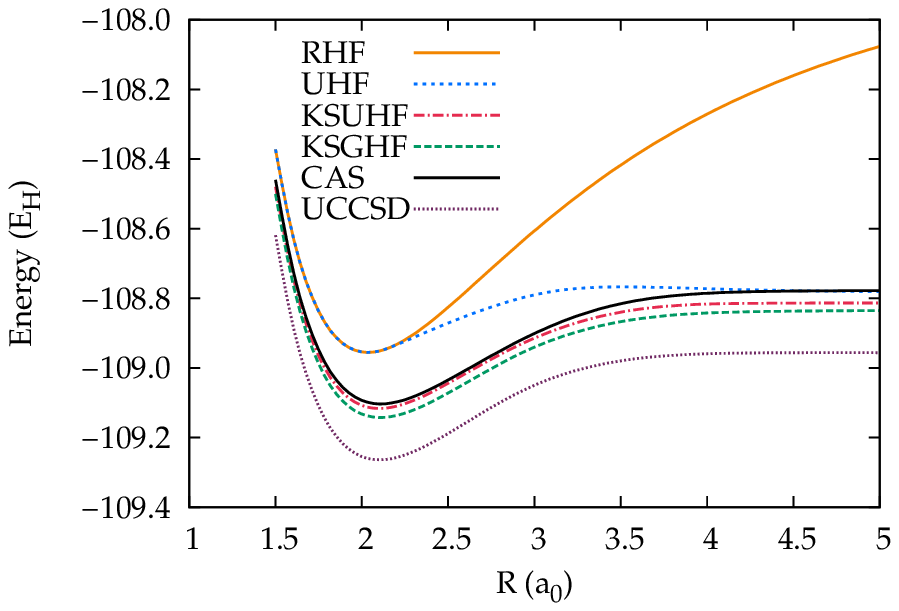}
\caption{Effects of simultaneous spin and complex conjugation projection on the potential energy curves for N$_2$ dissociation in the cc-pVDZ basis set, compared to the CASSCF(10/8) reference.
\label{Fig:N2Diss3}}
\end{figure}

While PHF yields excellent results for the dissociation of H$_2$, it misses some of the effects of dynamical correlation, which we illustrate by considering the dissociation of N$_2$, as shown in Figs. \ref{Fig:N2Diss1} - \ref{Fig:N2Diss3}.  The projected HF methods generally go to a dissociation limit slightly below that of UHF.  Near equilibrium, breaking and restoring complex conjugation symmetry accounts for most of the correlation available in CASSCF(10/8), though KRHF dissociates to a limit much too high.  Spin projection to give SUHF or SGHF is comparable to complex conjugation projection at dissocation, but generally inferior at equilibrium.  Combining the two in KSUHF and KSGHF gives energies below the (variational) CASSCF.  All of these curves, however, are far above the coupled cluster singles and doubles (CCSD) based on the UHF reference, even with this small basis set.  It is interesting to note that the KSGHF solution nearly parallels the UCCSD solution; the non-parallelity error (maximum deviation $-$ minimum deviation) is 8 kcal/mol.

\begin{figure}
\includegraphics[width=\columnwidth]{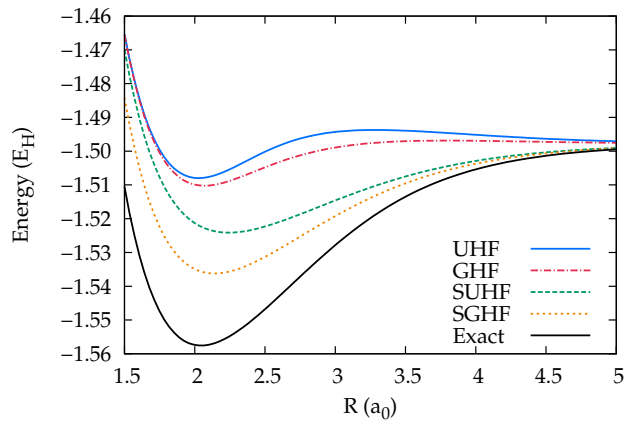}
\caption{Potential energy curve for the dissociation of equilateral H$_3$ in the cc-pVDZ basis set.
\label{Fig:H3Diss}}
\end{figure}

A second illustration of the missing dynamical correlation is provided by considering the dissociation of H$_3$.  Here, we arrange the three atoms on the corners of an equilateral triangle and expand the triangle symmetrically.  The Hartree--Fock ground state in this case is non-collinear (\textit{i.e.} is an eigenfunction of neither $\hat{S}^2$ nor $\hat{S}_z$).  Figure \ref{Fig:H3Diss} shows the dissociation, projecting onto $s=1/2$.  Even in this simple three-electron system, the spin projection clearly misses a significant portion of the total correlation.

\begin{figure}
\includegraphics[width=\columnwidth]{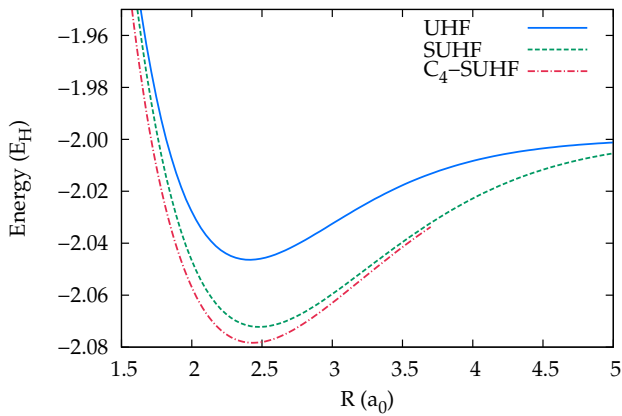}
\caption{Potential energy curve for the dissociation of square H$_4$ in the uncontracted STO-6G basis.  C$_4$-SUHF denotes the combination of spin restoration on a collinear deformed determinant and restoration of C$_4$ rotational symmetry.
\label{Fig:H4Diss}}
\end{figure}

An application of spatial symmetry restoration is provided in Fig. \ref{Fig:H4Diss}, where the isotropic expansion of a square of hydrogen atoms is shown.  We have performed these calculations with an uncontracted STO-6G basis set, and we have restored point group symmetry in the framework of the abelian $C_4$ subgroup of the full $D_{4h}$ symmetric group.  In particular, the lowest energy C$_4$-SUHF solution corresponds to B symmetry.  One can observe that symmetry breaking and restoration of the point group of the molecule provides additional correlation energy to SUHF, though the improvements relative to SUHF decrease as the system is pulled apart.

\begin{figure}
\includegraphics[width=\columnwidth]{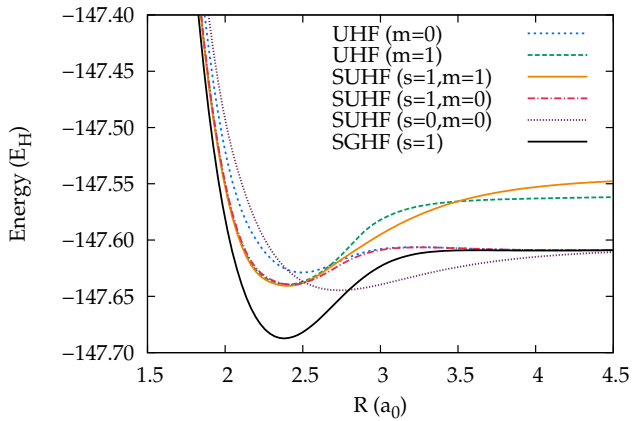}
\caption{Potential energy curve for O$_2$ dissociation in the STO-3G basis set.
\label{Fig:O2Diss}}
\end{figure}

Unlike standard Hartree--Fock and most previous EHF calculations, we can calculate all the $m$ components for a state with spin $s$.  In the case of SGHF, all these states are degenerate, as they should be.  This degeneracy is lost with SUHF, but a SUHF state can be found for each pair of quantum numbers ($s$, $m$).  Figure \ref{Fig:O2Diss} shows the dissociation of O$_2$ to the singlet and the $m = 0$ and $m = \pm 1$ triplets, computed with a minimal (STO-3G) basis.  The UHF dissociation on the triplet (more precisely, on the $m = 1$) curve is qualitatively right at equilibrium but has a bump before breaking spatial symmetry and dissociating from above.  The lowest energy UHF solution at dissociation has $m = 0$.  We have two different SUHF curves for $s=1$.  The $m = 1$ curve is qualitatively reasonable near equilibrium and has no bump, but goes to a limit above the UHF curve for the same $m$.  That UHF is below SUHF in this case is presumably because the UHF is contaminated by the quintet solution, which has lower energy at dissociation than does the triplet.  The $m = 0$ follows the UHF ``singlet'' and is more correct at dissociation.  Note that while SGHF has the attractive feature of making all components for a given spin state degenerate, finding SGHF solutions is technically demanding, in part because GHF solutions are generally unavailable but also because there are sometimes multiple SGHF solutions similar in energy.

\begin{figure}
\includegraphics[width=\columnwidth]{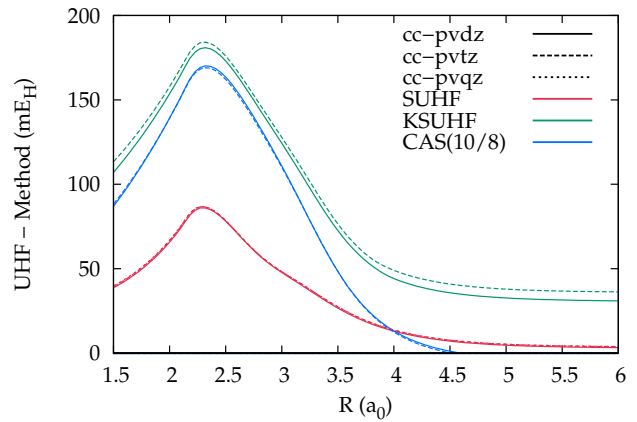}
\caption{Energy relative to UHF for the dissociation of N$_2$.  Notice that both CAS(10/8) and SUHF show essentially no dependence on the size of the basis, while KSUHF has some small basis set sensitivity; cc-pvqz results are only shown for SUHF.
\label{Fig:N2Corr}}
\end{figure}

Another interesting aspect that has not been thoroughly studied in the literature before is the basis set dependence of the correlation energy recovered by projection methods.  In Fig. \ref{Fig:N2Corr}, we show the difference of the SUHF, KSUHF, and CASSCF(10,8) energies with respect to UHF as a function of the internuclear separation with basis sets of increasing size.  As it is evident from the figure, the correlation energy recovered is almost independent of the basis set size for both SUHF and CASSCF.  This suggests that the correlations we recover are mostly static in nature.

\subsection{Size Consistency}
What is not so clear from the dissociation curves we have shown is that projected Hartree--Fock is not usually size consistent.  The magnitude of the error may be fairly large.  For example, the KSUHF dissociation limit of N$_2$ is roughly 27 mH (17 kcal/mol) higher than twice the KSUHF energy of the nitrogen atom, with comparable errors for KUHF and SUHF.  The origin of this size inconsistency is simply that SUHF on N$_2$ demands that the entire system be a singlet, but does not independently project any particular spin component onto the individual nitrogen atoms, which should each be a quartet.

In principle, a spin-projected Hartree--Fock method could be made size consistent with the aid of local projection operators.  That is, suppose we wish to dissociate a molecule to fragments $A$ and $B$.  Writing $|\Phi_A\rangle$ ($|\Phi_B\rangle$) for the broken symmetry reference for fragment $A$ (fragment $B$), we could imagine writing
\begin{equation}
\begin{split}
|\Psi\rangle
  &= \hat{P}_{AB} [(\hat{P}_A |\Phi_A\rangle) \otimes (\hat{P}_B |\Phi_B\rangle)]
\\
  &= \hat{P}_{AB} \hat{P}_A  \hat{P}_B |\Phi_A \Phi_B\rangle
\end{split}
\end{equation}
where $\hat{P}_A$ and $\hat{P}_B$ are projection operators which restore symmetry only on the individual (noninteracting) fragments, and $\hat{P}_{AB}$ subsequently restores symmetry on the whole system.  We could, for example, dissociate O$_2$ to two triplet oxygen atoms with the aid of $\hat{P}_A$ and $\hat{P}_B$, and put the whole system into a spin triplet via $\hat{P}_{AB}$.  This wavefunction is explicitly size consistent.

While formally this looks like projection operators acting on a broken symmetry determinant to restore symmetry, this approach lies outside the conventional PHF framework because we have relied on local projectors $\hat{P}_A$ and $\hat{P}_B$ in addition to a global projector $\hat{P}_{AB}$.  Unfortunately, it is far from clear how best to define such local projectors except for noninteracting subsystems.  The relation between the method sketched above and PHF is closely akin to the relation between AGP (which is not size consistent) and the size consistent Antisymmetrized Power of Strongly Orthogonal Geminals\cite{Surjan99} wherein one creates orthogonal subsets of orbitals and introduces geminals which pair orbitals only within a subset.

\subsection{The Curse of the Thermodynamic Limit}
We have already seen that projected HF is not size consistent.  Neither is it size extensive.  We demonstrate this by examining equally spaced rings of hydrogen atoms with an internuclear distance of 1.8 bohr.  As the number of atoms in the ring gets large, the ring approaches a periodic chain.  All calculations are done in the STO-6G basis set and for an overall spin singlet state.

\begin{figure*}
\includegraphics[width=\columnwidth]{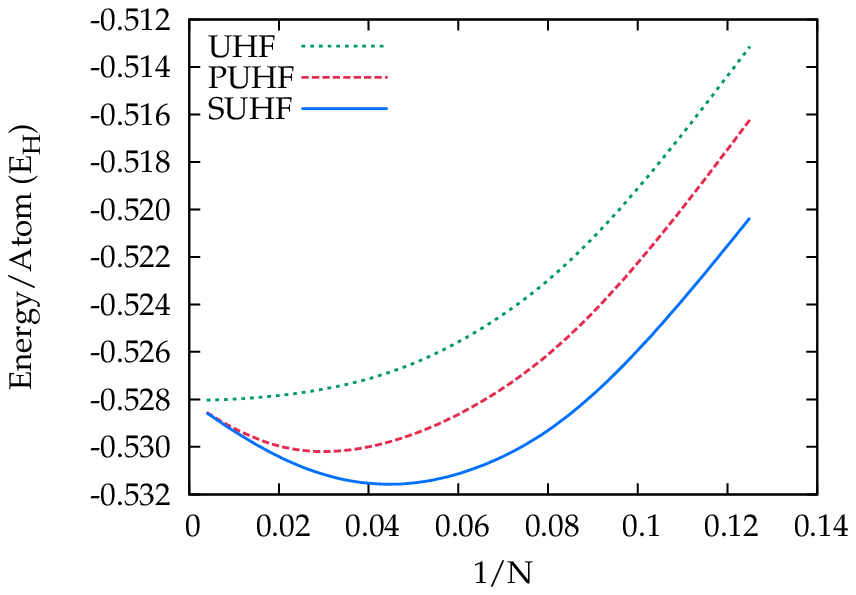}
\hfill
\includegraphics[width=\columnwidth]{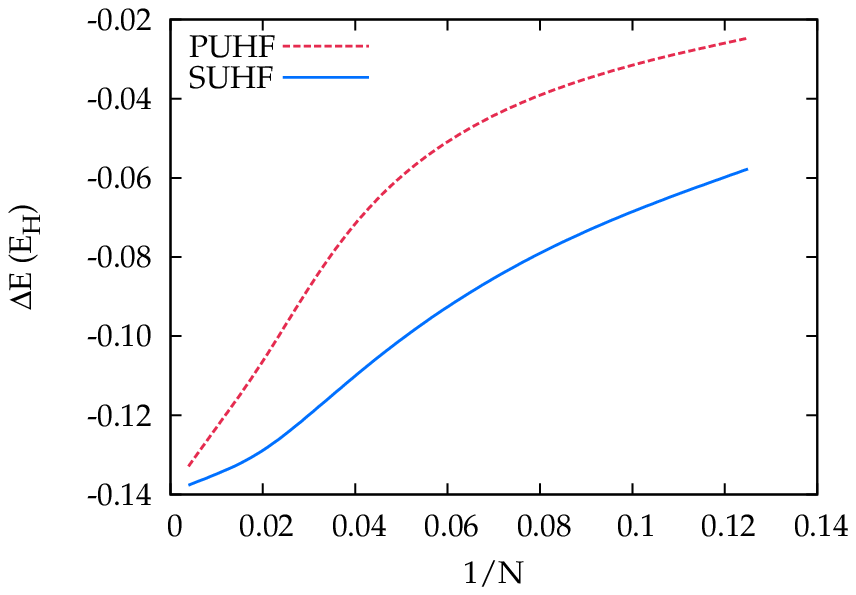}
\caption{Left Panel: Energy per atom in an equally spaced hydrogen ring.  Right Panel: Energetic improvement of spin projection over UHF.  Both panels use one over the number of atoms for the abscissa; the thermodynamic limit is $1/N \to 0$. Color online. 
\label{Fig:HRing}}
\end{figure*}

In Fig. \ref{Fig:HRing} we show the energy per atom as a function of the ring size.  While for small rings UHF, PUHF, and SUHF all differ noticeably, the three curves converge to the same energy per atom as the rings become large.  Figure \ref{Fig:HRing} also shows that as the number of particles becomes large, both PUHF and SUHF give the same improvement in total energy over UHF.  In other words, both PUHF and SUHF provide the same correlation energy relative to UHF, (about 140 mH), but the physically meaningful correlation energy per particle vanishes for large $N$.  This observation was made by Mayer and Kert\'esz\cite{Mayer76} and by Casta\~no and Karadakov\cite{Castano86} in a PPP model of finite polyene chains, and our results show the behavior they described.  In the limit $N \to \infty$, where $N$ is the number of particles, restoring spin symmetry has no effect on the energy per particle.  This is what we mean when we say that projected Hartree--Fock is not size extensive.

The rationale behind the failure of projected methods to go beyond broken symmetry mean-field theory in the thermodynamic limit is as follows.  For finite systems, the energy lowering due to projection comes about from the fact that the matrix elements $\langle \Phi | \hat{H} \hat{R} | \Phi \rangle$ and $\langle \Phi | \hat{R} | \Phi \rangle$ are non-zero, where $\hat{R}$ are rotation matrices, as explained in Section \ref{Sec:Theory}.  As the system size increases, those matrix elements tend to zero.  If they are zero, then the Hamiltonian becomes diagonal in the basis of the rotated states, and thus there is no chance for energy lowering to occur.  In other words, the Goldstone  manifold (defined in Ref. \onlinecite{Scuseria11}) remains degenerate but becomes non-interacting.

\subsection{Singlet-Triplet Splittings}
Spin splittings (the difference in energy between the lowest energy states of various spins $s$) are important quantities in both organic and inorganic chemistry.  They are connected to magnetic coupling constants in the Heisenberg Hamiltonian, and can be unfortunately challenging to evaluate with simple mean-field methods.  Density functional theory is well known to have difficulties with spin splittings, and unrestricted Hartree--Fock (UHF) also often fails, particularly when spin contamination is large.\cite{Davidson04}

One can attempt to remedy the failings of UHF by projecting the UHF wavefunction onto spin eigenfunctions (that is, to use a PAV approach), but in many cases this projected UHF is still inadequate.  The question we wish to address here is whether variation after projection in our PHF scheme offers more accurate results.  We will focus on singlet-triplet splittings, as they are the most commonly studied.

In Table \ref{Tab:Splittings} we show the computed singlet-triplet splittings for a variety of small diatomic molecules,\cite{Huber} as well as methylene (CH$_2$),\cite{CH2Exp} trimethylmethane (TMM),\cite{TMMExp} and \textit{o}-, \textit{m}-, and \textit{p}-benzyne.\cite{Benzyneexp}  All calculations use the cc-pVTZ basis set, and we define the singlet-triplet splittings as simply
\begin{equation}
\Delta E_{ST} = E_T - E_S
\label{Est}
\end{equation}
where $E_T$ and $E_S$ are the energies of the triplet and the singlet states, respectively.  We have included results using CUHF, taken from Ref. \onlinecite{Tsuchimochi11}. In CUHF, a triplet state is obtained by optimizing the HF determinant subject to the constraint that it remains a spin eigenfunction. For singlet states, CUHF allows symmetry breaking only between two orbitals. PCUHF corresponds to spin-projection on CUHF performed with a PAV approach.

\begin{table*}
\caption{Singlet triplet splittings (kcal/mol) from several methods.  ``TMM'' refers to trimethylmethane, while ME is the mean error (theory $-$ experiment) and MAE is the mean absolute error.
\label{Tab:Splittings}}
\begin{ruledtabular}
\begin{tabular}{lcccccccc}
Molecule            &  UHF  &  KUHF  &  CUHF  &  PUHF  & PCUHF  &  SUHF  & KSUHF  &  Expt  \\
\hline
NH                  &  19.4 &  18.6  &  21.0  &  38.2  &  41.5  &  33.6  &  31.6  &  39.0 \\
OH$^+$              &  25.9 &  25.0  &  27.4  &  50.9  &  54.1  &  45.8  &  43.4  &  50.6 \\
O$_2$               &  15.8 &  14.6  &  16.1  &  26.8  &  30.6  &  20.6  &  24.2  &  22.6 \\
NF                  &  19.7 &  18.6  &  20.6  &  38.1  &  40.8  &  32.3  &  31.0  &  34.3 \\
CH$_2$              &  16.9 &  15.9  &  15.4  &  18.9  &  15.7  &  15.6  &  14.0  &   9.4 \\
TMM                 &  23.7 &  15.6  &   7.5  &  28.6  &  10.5  &  19.1  &  18.0  &  17.7 \\
\textit{o}-benzyne  & -15.8 & -12.5  &  -9.8  & -42.6  & -24.5  & -51.4  & -48.7  & -38.0 \\
\textit{m}-benzyne  &  28.3 &  13.0  &  12.8  &  45.0  &  12.8  &   2.2  &  -9.1  & -20.6 \\
\textit{p}-benzyne  & -10.1 &  -5.2  &   0.1  & -24.0  &  -0.5  & -28.2  & -22.9  &  -3.5 \\
\hline
ME                  &   1.3 &  -1.0  &   0.0  &   7.6  &   7.7  &  -2.5  &  -3.4  \\
MAE                 &  17.5 &  15.5  &  15.9  &  13.4  &   9.3  &   9.2  &   7.4  \\
\end{tabular}
\end{ruledtabular}
\end{table*}

For the small diatomic molecules, both SUHF and PUHF provide excellent agreement with experiment.  As the systems become larger (but not too large; see above) SUHF and PUHF differ more, and SUHF is more accurate.  Generally, we observe that spin projection yields meaningful improvements when the spin contamination in the UHF is large only for the singlet state, but has less to offer when the UHF has significant spin contamination in both the singlet and triplet states.  This is demonstrated by the benzynes, where SUHF, though more accurate than UHF or PUHF, is still far from experiment.  Complex conjugation restoration has, as one might expect, much smaller effects on the calculated singlet-triplet splittings.  We should point out that dynamic correlation can have large effects on calculated singlet-triplet splittings which cannot be captured by SUHF and SGHF as these methods account primarily for static correlation.

Note that semi-empirical corrections to calculated singlet-triplet splittings with spin contaminated wavefunctions such as UHF and KUHF can be extracted as, for example,\cite{Noodleman,Yamaguchi,Davidson}
\begin{equation}
\Delta E_{ST} = 2 \frac{E_T - E_S}{\langle \hat{S}^2 \rangle_T - \langle \hat{S}^2 \rangle_S}
\label{Estcheating}
\end{equation}
where $E_T$ and $E_S$ are the energies of the spin contaminated triplet and singlet determinants and $\langle \hat{S}^2 \rangle_T$ and $\langle \hat{S}^2 \rangle_S$ are the respective expectation values of $\hat{S}^2$.  If the wavefunctions used are spin eigenfunctions, Eqn. \ref{Estcheating} reduces to Eqn. \ref{Est}.  While the results for spin contaminated states are closer to experiment, the prescription is neither unique nor first principles.  Applying this correction to KUHF leads to a MAE of 10.1 kcal/mol, still worse than KSUHF.

\section{Conclusions
 \label{Sec:Conclusions}}
Symmetries and good quantum numbers are of critical importance in many finite quantum systems.  Unfortunately, approximate variational solutions to the Schr\"odinger equation need not respect the same symmetries as does the exact solution.  Forcing them to do so reduces the variational flexibility of the model, which is not ideal.  On the other hand, symmetry breaking from mean-field theories generally indicates the failure of the mean-field approximation and the emergence of strong correlation.  One can take advantage of this symmetry breaking to obtain energetically reasonable broken symmetry mean-field wavefunctions, though the wavefunctions obtained are of poor quality.  By using projection operators, one can restore the broken symmetries and recover the associated quantum numbers while obtaining strongly correlated wavefunctions. Moreover, this can be accomplished without leaving the independent particle picture that allows one to easily grasp the physics in the wavefunction.

While self-consistently restoring the symmetry of mean-field wavefunctions is formally attractive, the practical realization of it is not trivial. On the other hand, a projection-after-variation approach is simple, but may lead to severe problems with the projected wavefunction. A self-consistent approach also allows one to break symmetry in a deliberate way, that is, one can find a broken symmetry reference determinant even in cases when the standard HF method leads to a symmetry-adapted state.

Our earlier work showed how to optimize the energy of the projected wavefunction when the broken symmetry reference state was a Hartree--Fock--Bogoliubov determinant that broke particle number symmetry.  Here, we do the same when the reference determinant conserves electron number.  There are two reasons to present this new set of PHF equations.  First, due to the historical significance of the PHF problem in quantum chemistry.  Second, the PQT equations as presented before lead to indeterminancies for systems whose deformed reference state has exact zero occupations, as is the case when particle number symmetry is not broken. The pairing interaction in PQT assigns a non-zero occupation to every natural orbital, but in practice the occupation numbers may become sufficiently small in a large enough basis that the equations cannot be accurately solved in double precision arithmetic.  The ideas presented here can be extended to remedy this deficiency.

The PHF formulation here presented is easy to code and numerically robust.  The resulting projected Hartree--Fock scheme is conceptually simple and computationally affordable, and generally leads to good quality multireference wavefunctions.  Size consistency and size extensivity remain problems, but otherwise projected Hartree--Fock offers a relatively black box treatment of strong correlation while staying within the mean-field picture without sacrificing good quantum numbers.

\section{Acknowledgments}
This work is supported by the National Science Foundation under CHE-0807194 and CHE-1110884, the Welch Foundation (C-0036), and Los Alamos National Labs (Subcontract 81277-001-10).

\appendix
\section{The PHF Effective Hamiltonian}
Here, we provide the explicit expressions required to evaluate the matrix elements of the effective Fock matrix.  The PHF energy expression can be conveniently written as
\begin{equation}
E = \int \mathrm{d}\Omega \, y(\Omega) \, \left\{ \mathrm{Tr}[\bm{h}  \, \bm{\rho}_{\Omega}] + \frac{1}{2} \mathrm{Tr}[\mathbf{G}_\Omega \, \bm{\rho}_\Omega]\right\}
\end{equation}
where $\bm{h}$ is the matrix of one-electron integrals and $\mathbf{G}_\Omega$ is 
\begin{equation}
(\mathbf{G}_\Omega)_{ik} = \sum_{jl} \langle ij || kl \rangle (\bm{\rho}_\Omega)_{lj}.
\end{equation}
The effective Fock matrix is defined in terms of derivatives of the PHF energy expression with respect to matrix elements of the density matrix (of the deformed state) as
\begin{equation}
\mathcal{F}_{kl} = \frac{\partial\hfill }{\partial \rho_{lk}} E[\bm{\rho}].
\label{defF}
\end{equation}
$\bm{\mathcal{F}}$ is Hermitian by construction, that is,
\begin{equation}
\mathcal{F}^\ast_{lk} = \frac{\partial\hfill }{\partial \rho^{\ast}_{kl}} E[\bm{\rho}] =  
\frac{\partial\hfill }{\partial \rho_{lk}} E[\bm{\rho}] = \mathcal{F}_{kl}.
\end{equation}

We follow Sheikh and Ring\cite{Sheikh2000} in writing the derivatives of the function $y(\Omega)$ in the form
\begin{equation}
\frac{\partial\hfill}{\partial \rho_{kl}} \, y(\Omega) =  y(\Omega) \, (\mathbf{Y}_\Omega)_{lk}.
\end{equation}
Using this, we can write the effective Fock matrix as
\begin{equation}
\begin{split}
\mathcal{F}_{ij} = \int \mathrm{d}\Omega \, y(\Omega)
    &  \Big\{(\mathbf{Y}_\Omega)_{ij} \left[\mathrm{Tr}(\bm{h} \,  \bm{\rho}_\Omega) + \frac{1}{2} \mathrm{Tr}(\mathbf{G}_\Omega \, \bm{\rho}_\Omega)\right]
\\
    & + \sum_{kl} (\bm{h} + \mathbf{G}_\Omega)_{lk} \frac{\partial \hfill}{\partial \rho_{ji}} (\bm{\rho}_\Omega)_{kl} \Big \}.
\end{split}
\end{equation}

The derivation of the matrices $\mathbf{Y}_\Omega$ and $\bm{\mathcal{F}}$ is straightforward but cumbersome.  Our final expression for $\mathbf{Y}_\Omega$ is
\begin{equation}
\mathbf{Y}_\Omega = \bm{\mathcal{Y}}_\Omega - \int \mathrm{d}\Omega' \, y(\Omega') \, \bm{\mathcal{Y}}_{\Omega'}
\label{DefY}
\end{equation}
where
\begin{equation}
\bm{\mathcal{Y}}_\Omega 
  = \mathbf{R}_\Omega \, \begin{pmatrix} \bm{\rho}_{pp}  \\ \bm{\rho}_{qp} \end{pmatrix}  \, \mathbf{N}_\Omega
 \, + \,
    \mathbf{N}_\Omega \, \begin{pmatrix} \bm{\rho}_{pp} &  \bm{\rho}_{pq} \end{pmatrix} \, \mathbf{R}_\Omega.
\label{defYOmega}
\end{equation}

If we define $\mathbf{F}_\Omega = \bm{h} + \mathbf{G}_\Omega$, we can write $\bm{\mathcal{F}}$ as
\begin{equation}
\bm{\mathcal{F}} = \int \mathrm{d}\Omega \, y(\Omega) \, \bm{\mathcal{F}}_\Omega
\end{equation}
where
\begin{align}
\bm{\mathcal{F}}_\Omega 
  &= \frac{1}{2} \mathbf{Y}_\Omega  \mathrm{Tr}[(\bm{h} + \mathbf{F}_\Omega) \bm{\rho}_\Omega]
\nonumber
\\
  &+ \mathbf{N}_\Omega \begin{pmatrix} \bm{\rho}_{pp} &  \bm{\rho}_{pq} \end{pmatrix} \mathbf{F}_\Omega (\bm{1} - \bm{\rho}_\Omega) \mathbf{R}_\Omega
\nonumber
\\
  &+ (\bm{1} - \bm{\rho}_\Omega) \mathbf{F}_\Omega \mathbf{R}_\Omega \begin{pmatrix} \bm{\rho}_{pp} \\ \bm{\rho}_{qp} \end{pmatrix} \mathbf{N}_{\Omega}.
\label{defFOmega}
\end{align}

Note that since $\mathbf{N}_\Omega$ is an $N \times N$ matrix, both $\bm{\mathcal{Y}}_\Omega$ and parts of $\bm{\mathcal{F}}_\Omega$ are given as a sum of an $M \times N$ and an $N \times M$ matrix.  These matrices are correctly understood as $M \times M$, padded to the right or below by zeroes.  The fact that $(\bm{\mathcal{F}}_\Omega)_{qq}$ and $(\bm{\mathcal{Y}}_\Omega)_{qq}$ vanish is simply a consequence of the fact that the energy is independent of $\bm{\rho}_{qq}$.

In the molecular orbital basis of $|\Phi\rangle$ in which we work in practice, we have
\begin{equation}
\bm{\rho} = \begin{pmatrix} \mathbf{1}  & \mathbf{0}  \\ \mathbf{0}  & \mathbf{0}  \end{pmatrix}.
\end{equation}
The subscripts $p$ and $q$ here will be replaced by $o$ and $v$ to emphasize that they denote occupied and virtual orbitals, respectively.  Writing
\begin{equation}
\mathbf{R}_\Omega = \begin{pmatrix} \mathbf{R}_{oo}  &  \mathbf{R}_{ov}  \\  \mathbf{R}_{vo}   &  \mathbf{R}_{vv}  \end{pmatrix},
\end{equation}
where we have suppressed the dependence on $\Omega$ for brevity, we obtain
\begin{equation}
\mathbf{N}_\Omega = \mathbf{R}_{oo}^{-1}
\end{equation}
where $\mathbf{R}_{oo}^{-1}$ is the inverse of $\mathbf{R}_{oo}$.  The transition density matrices are then
\begin{equation}
\bm{\rho}_\Omega = \begin{pmatrix} \mathbf{1}  &  \mathbf{0}  \\  \mathbf{R}_{vo} \mathbf{R}_{oo}^{-1}  & \mathbf{0}  \end{pmatrix}.
\end{equation}
One finds that
\begin{equation}
\bm{\mathcal{Y}}_\Omega = \begin{pmatrix}  \mathbf{2}  &  \mathbf{R}_{oo}^{-1} \mathbf{R}_{ov}  \\  \mathbf{R}_{vo} \mathbf{R}_{oo}^{-1}  &  \mathbf{0}  \end{pmatrix}.
\end{equation}

Finally, we have
\begin{equation}
\bm{\mathcal{F}}_\Omega = \frac{1}{2} \mathbf{Y}_\Omega  \mathrm{Tr}\left[(\bm{h} + \mathbf{F}_\Omega) \bm{\rho}_\Omega\right] + \tilde{\bm{\mathcal{F}}}_\Omega
\end{equation}
where the individual blocks of $ \tilde{\bm{\mathcal{F}}}_\Omega$ are
\begin{subequations}
\begin{align}
(\tilde{\bm{\mathcal{F}}}_\Omega)_{oo} &=  \mathbf{0},
\\
(\tilde{\bm{\mathcal{F}}}_\Omega)_{ov} &=  \mathbf{R}_{oo}^{-1} \mathbf{F}_{ov} \left[ \mathbf{R}_{vv} - \mathbf{R}_{vo} \mathbf{R}_{oo}^{-1} \mathbf{R}_{ov}\right],
\\
(\tilde{\bm{\mathcal{F}}}_\Omega)_{vo} &= \mathbf{F}_{vo} -  \mathbf{R}_{vo} \mathbf{R}_{oo}^{-1} \mathbf{F}_{oo} 
\\
                                     &+ \left[ \mathbf{F}_{vv} - \mathbf{R}_{vo} \mathbf{R}_{oo}^{-1} \mathbf{F}_{ov}\right]  \mathbf{R}_{vo} \mathbf{R}_{oo}^{-1},
\nonumber
\\
(\tilde{\bm{\mathcal{F}}}_\Omega)_{vv} &= \mathbf{0}.
\end{align}
\end{subequations}
Using Eqn. \ref{DefY} and the fact that $y(\Omega)$ integrates to unity, it is clear that the occupied-occupied and virtual-virtual blocks of $\mathbf{Y}_\Omega$ vanish, and thus so do these blocks of $\bm{\mathcal{F}}_\Omega$.

\bibliography{PHF}

\end{document}